\documentstyle[aps,preprint,prl]{revtex}
\tightenlines

\begin{document}
\bibliographystyle{prsty}
\draft

\title{{\it Ab initio} phonon dispersion curves and \\
interatomic force constants of barium titanate}
\author{Ph. Ghosez, X. Gonze and J.-P. Michenaud}
\address{Unit\'e de Physico-Chimie et de Physique des Mat\'eriaux,
Universit\'e Catholique de Louvain,\\
1 Place Croix du Sud, B-1348 Louvain-la-Neuve, Belgium}
\date{\today}
\maketitle
\begin{abstract}
The phonon dispersion curves of cubic BaTiO$_3$ have been computed within a
first-principles approach and the results compared to the experimental data. The curves
obtained are very similar to those reported for KNbO$_3$ by Yu and Krakauer [Phys. Rev.
Lett. {\bf 74}, 4067 (1995)]. They reveal that correlated atomic displacements along $<$100$>$
chains are at the origin of the ferroelectric instability. A simplified model illustrates
that spontaneous collective displacements will occur when a dozen of aligned atoms are
coupled. The longitudinal interatomic force constant between nearest neighbour Ti and O
atoms is relatively weak in comparison to that between Ti atoms in adjacent cells. The small
coupling between Ti and O displacements seems however necessary to reproduce a
ferroelectric instability.
\end{abstract}

\setcounter{page}{1}
\newpage

\section{introduction}
Barium titanate (BaTiO$_3$) is well known to exhibit three ferroelectric phase transitions:
stable at high temperature in a perovskite cubic phase, its structure becomes successively
tetragonal, orthorhombic and finally rhombohedral as the temperature goes down. Many
works investigated and discussed the origin of its ferroelectric phase
transitions~\cite{Lines77}. Amongst them, the most gratifying contribution  was probably due
to Cochran~\cite{Cochran60} who associated the ferroelectric instability with the softening
of a transverse optic phonon, and emphazised the connection between the structural
instability and the lattice dynamics. 

Consequently to Cochran's work, a large amount of experiments have been performed  in
order to confirm the existence of a soft ferroelectric mode in BaTiO$_3$. They include
infra-red~\cite{Luspin80,IR} and Raman~\cite{Raman} measurements of the $\Gamma$
phonon modes as well as various neutron diffraction
data~\cite{Shirane67,Yamada69,Shirane70,Harada71,Bouillot79,Jannot84}. These
experiments focused on the temperature behaviour of the soft phonon and were mainly
concerned by the low frequency modes: only a few results are available for the higher
energy vibrations.

Simultaneously, theoretical phonon dispersion curves were deduced from fits to the
experimental data using different shell models. Let us mention the pseudo-ionic model
developped by Gnininvi and Bouillot~\cite{Gnininvi77} or the rigid-shell model used by Jannot
{\it et al.}~\cite{Jannot84}. These models were however not particularly suited to describe
ABO$_3$ crystals, like BaTiO$_3$. Migoni, Bilz and B\"auerle~\cite{Migoni76} pointed out that
the behaviour of the ferroelectric soft mode in the oxidic perovskites originates from the
unusual anisotropic polarizability of the oxygen that, in turn, may be connected to
hybridization between O-2p and B-d states. Consequently,  a more sophisticated
``polarizability model''~\cite{Bilz87+} was introduced in order to include the specific physical
features of ABO$_3$ compounds. The application of this model to BaTiO$_3$ was reported by
Khatib {\it et al.}~\cite{Khatib89}. In their work, they obtained a full phonon band structure
and  investigated the temperature behaviour of the ferroelectric soft mode. However, their
interesting results still remained at a semi-empirical level.

Since a few years, theoretical advances have enabled one to determine the phonon
dispersion curves of solids within a truly first-principle approach. Recently, {\it ab initio}
results have been reported for KNbO$_3$~\cite{Yu95} and SrTiO$_3$~\cite{LaSota96}.
Similarly, in this paper,  we investigate the lattice dynamics of BaTiO$_3$. 

Our phonon frequencies compare well with the experimental data available. Moreover, our
dispersion curves are very similar to those reported for KNbO$_3$~\cite{Yu95}, a
perovskite material presenting the same sequence of phase transition as BaTiO$_3$. In
particular, the wave vector of the unstable phonon modes remains located in three $(110)$
slab region of the Brillouin zone, emphasizing a chain-structure instability in real
space~\cite{Yu95}.  This behaviour is illustrated with a simplified model. The range and
anisotropy of the interatomic force constants is also discussed.

\section{Technicalities}

Our calculations have been performed in the general framework of the density functional
formalism~\cite{DFT}. The exchange-correlation energy functional was evaluated within the
local density approximation, using a polynomial parametrization of
Ceperley-Alder~\cite{Ceperley80} homogeneous electron gas data.

We used a plane-wave pseudopotential approach. The all-electron potentials were replaced
by the same extended norm conserving highly transferable pseudopotentials as in
Ref.~\cite{Ghosez94}. We have considered 5s, 5p and 6s as valence states to build the Ba
pseudopotential, 3s, 3p, 3d and 4s valence states for the Ti pseudopotential and 2s and
2p valence states for the O pseudopotential. The electronic wave function was expanded in
plane waves up to a kinetic energy cutoff of 45 Hartree (about 6200 plane waves). Integrals
over the Brillouin zone were replaced by a sum on a mesh of 6X6X6 special
$k$-points~\cite{Monkhorst76+}.

The dynamical matrix, Born effective charges and dielectric tensor were computed
within a variational formulation~\cite{Gonze92} of the density functional perturbation
theory~\cite{DFPT}.   First, calculations were carried out to determine the dynamical
matrix on different meshes of $q$-points.  Then, an interpolation was performed following
the scheme proposed in Ref. \cite{Giannozzi91,Gonze94}. This technique takes properly into
account the long-range behaviour of the dipole-dipole interaction which is separated
from the remaining short-range forces owing to the knowledge of the Born effective
charges and the optical dielectric tensor. 

An insight into the convergence reached on the phonon dispersion curves is reported in
Fig.~\ref{Fig.GR}. The frequencies deduced from the dynamical matrix at
$q=(.125, .125, .125)$ and $q=(.375, .375, .375)$ are compared to those extrapolated from
two different meshes of $q$-points: the first mesh (M1) includes $\Gamma$ (.0, .0, .0), X (.5,
.0, .0), M (.5, .5, .0) and R (.5, .5, .5) points; the second mesh (M2) is the cubic mesh M1 to
which the $\Lambda$ (.25, .25, .25) point was added.  We obtain a very good convergence with
the M2 mesh. 

\section{The phonon dispersion curves}

Our calculations are performed at the experimental lattice parameter of 4.00 \AA. This
choice facilitates the comparison with the experimental data. Some indications on the
volume dependence of the phonon frequencies can be found in Ref.~\cite{Ghosez96+}, where
the frequencies of the $\Gamma$ phonons at different lattice constants have been
compared.

First, we present in Table~\ref{Table.Z} the Born effective charges and the optical
dielectric constant~\cite{Comment2}, important ingredients of the present study since their
knowledge allows to identify the long-range part of the interatomic force constants and
makes the interpolation of phonon frequencies tractable.  For Ba and Ti atoms, the effective
charge tensor is isotropic. For O, the two independent components of the tensor correspond
respectively to a displacement of the atom parallel ($Z^*_{O \parallel}$) and perpendicular
($Z^*_{O \perp}$) to the Ti-O bond. $Z^*_{Ti}$ ($+7.32$) and $Z^*_{O \parallel}$ ($-5.78$) are
anomalously large~\cite{Ghosez94,Axe67,Zhong94,Ghosez95+} with respect to the nominal
ionic charges (+4 for Ti and $-2$ for O). This surprising phenomenon was explained recently
in connection within dynamic changes of hybridization between O-2p and Ti-3d
orbitals~\cite{Ghosez95+,Posternak94}. This specific feature is at the origin of a large
destabilizing dipole-dipole interaction, connected to the anomalous mode effective charge
\cite{Ghosez96+,Zhong94} and the instability of the ferroelectric mode~\cite{Ghosez96+}. 

Our computed optical dielectric constant (6.75) largely overestimates the experimental
value (5.40)~\cite{Burns82}, as usual within the local density approximation. This
discrepancy will essentially affect the position of the highest longitudinal optic mode:
when replacing the theoretical dielectric constant by the experimental value, its frequency
at the $\Gamma$ point changes from 631 to 696 cm$^{-1}$. At the opposite, the frequencies of
the two other longitudinal modes at the $\Gamma$ point are affected by less than 2 cm$^{-1}$.

The calculated phonon dispersion curves are plotted along high symmetry directions in
Fig.~\ref{Fig.BS}.  The $\Gamma$-X, $\Gamma$-M and $\Gamma$-R lines are along the
$<$100$>$, $<$110$>$ and $<$111$>$ directions, respectively. The unstable modes associated to
a negative curvature of the energy hypersurface have imaginary phonon frequencies. The
frequencies at the high symmetry points are reported in Table~\ref{Table.w}.

Our result can be compared to the experimental
data~\cite{Luspin80,Shirane67,Yamada69,Shirane70,Harada71,Bouillot79,Jannot84}.
However, a complication arises from the fact that all the experimentally observed
vibrational excitations have a real frequency while the computed unstable modes are
obtained with an imaginary frequency. As the soft mode can be clearly identified by its
symmetry,  the associated experimental frequencies were removed from the comparison, for
clarity. In the low frequency region, the presence of this additional soft mode may have
slightly modified the frequency of the other modes. In spite of these difficulties we observe
a good correspondence between our theoretical frequencies and the experimental data,
specially for the acoustic modes for which a large variety of data are available. 

The ferroelectric phase transitions are driven by the unstable phonon modes. We are
therefore mainly concerned by the analysis of these specific phonons. Two transverse
optic modes are unstable at the $\Gamma$ point: they correspond to a displacement of the
Ti atom against the oxygen cage. The associated displacement eigenvector is equal to
[$\delta ({\rm Ba})= -0.002$, $\delta ({\rm Ti})=-0.096$, $\delta ({\rm O_1})=+0.158$, $\delta
({\rm O_2})=\delta ({\rm O_3})=+0.071$]~\cite{Comment1}. These two modes remain unstable
all along the $\Gamma$-X line, with very little dispersion. One of them stabilizes along the
$\Gamma$-M and X-M lines. Examination of the eigenvectors reveals that the unstable mode
at the M (.5, .5, .0) point is polarized along the $z$-direction: its displacement eigenvector is
equal to [$\delta ({\rm Ti_z})=-0.130$, $\delta ({\rm O_{1,z}})=+0.106$]~\cite{Comment1}. 
Both of the unstable modes become stable when deviating from the three $\Gamma$-X-M
planes to the R-point.  

These features were also observed for KNbO$_3$ \cite{Yu95} and point out a marked 2D
character of the instability in the Brillouin zone. This behaviour is more easily visualized in
Fig.~\ref{Fig.IM} where we show the frequency isosurface of the lowest unstable phonon
branch corresponding to $\omega=0$. The region of instability, $\omega^2({\bf q}) < 0$, lies
between three pairs of flat surfaces, that are parallel  to the faces of the Brillouin zone cube.
In other words, the unstable modes are contained in three perpendicular interpenetrating
slab-like regions of finite thickness containing the $\Gamma$ point. 

As highlighted by Yu and Krakauer \cite{Yu95}, this behaviour corresponds to chain
instabilities in real space. At the M-point, we have seen that there is a single unstable mode
polarized along the $z$-axis and dominated by the Ti$_z$ and O1$_z$ displacements. At this
wave vector ($q_z =0$), the Ti and O$_1$ atoms will be coherently displaced all along an
{\it infinite} $<$001$>$ chain. Going now from M to the R-point,  the coherency of the
displacement will gradually disappear and a {\it finite} length of correlation will be reached
for which the phonon becomes stable. The finite thickness of the slab region of instability
therefore corresponds to a minimum correlation length of the displacement required to
observe an unstable phonon mode. From Fig.~\ref{Fig.IM}, the length of the shortest unstable
chain can be estimated to $4 \; a_{cell} \; = \;$ 16 \AA \cite{Comment3}. Note finally, the small
dispersion of the unstable mode in the $\Gamma$-X-M plane suggests a small correlation of
the displacements between the different Ti--O chains.

\section{The interatomic force constants}

In cubic BaTiO$_3$, we will see that the single displacement of a particular atom never leads
to an instability: When one atom is displaced, a force is induced and brings it back in its initial
position (see Tables~\ref{Table.IFC} - \ref{Table.O-O}: the self-force on Ti and O is positive).
However, its atomic displacement will simultaneously induce forces on the other atoms. It is
only the additional displacement of some other atoms in this force field  that can lower the
total energy and produce an instability. The amplitude and the range of the interatomic force
constants (IFC) associated to this mechanism can be analysed in order to clarify
the chain instability pointed out in the previous Section. Moreover, the specific role of the
dipole-dipole interaction (DD) can be separated from that of the short-range forces (SR)  
following the scheme proposed in Ref. \cite{Gonze94}. Our conventions on the interatomic
force constants $C_{\alpha, \beta}(\kappa, \kappa')$ are such that the force
$F_{\alpha}(\kappa)$ induced on atom $\kappa$ by the displacement
$\Delta \tau_{\beta}(\kappa')$ of atom $\kappa'$ is given by: $F_{\alpha}(\kappa) = - C_{\alpha,
\beta}(\kappa, \kappa') \; . \; \Delta \tau_{\beta}(\kappa')$.

Let us first investigate the IFC with respect to a reference Ti atom along a Ti-O chain
(Table~\ref{Table.IFC}). As previously mentionned, we note that the self-force on the Ti
atom is large and positive (+0.15215 Ha/Bohr$^2$). We observe also that the longitudinal IFC
with the first neighbour O atom is surprisingly small (+0.00937 Ha/Bohr$^2$); moreover, it is
positive. The analysis of the DD and SR contributions points out that these characteristics are
the result of a destabilizing DD interaction, sufficiently large to compensate the SR forces. It
is this close compensation which allows the displacement of Ti against the O atoms. Another
insight on this balance of forces was already reported in Ref.~\cite{Ghosez96+}. Consequently
to the very small total IFC, the Ti and O displacements might be relatively decoupled.  

At the opposite, the DD forces induced on the next Ti atom are negative: they will combine
with the SR forces in order to produce sizeable coupling ($-$0.06721 Ha/Bohr$^2$). This
mechanism is at the origin of the chain correlation of the Ti atomic displacements.  By
contrast, the {\it transverse} force on the first Ti neighbour is very small and confirms the
small correlation of the displacements from chain to chain.

The decay of the Ti--Ti and O--O longitudinal IFC with the interatomic distance can also be
investigated. The results are reported in Table \ref{Table.Ti-Ti} and
\ref{Table.O-O}. It is seen that the longitudinal IFC are anisotropic: they propagate
essentially along the Ti--O chain. This appears clearly for the SR part. For O, the DD
contribution is also highly anisotropic due to the anisotropy of the Born effective charges.
The anisotropy of the IFC is inherent to the chain correlation previously mentionned in this
paper. 

\section{The chain-structure instability}

From the knowledge of the IFCs previously reported, we can also investigate the energy
surface of BaTiO$_3$ and illustrate the chain correlation highlighted in Section III.  Let us
consider that we have a bulk cubic crystal with the atoms frozen at their equilibrium position
$\tau_{\kappa_0}$. Then, we allow displacements of Ti and O atoms belonging to a [100] single
Ti--O chain of finite but increasing size. The total energy of this system will be given by:
$$
E({\tau_{\kappa}})= E({\tau_{\kappa_0}}) + 
                       \sum_{\kappa, \kappa'} \; C_{1,1}(\kappa, \kappa') \; 
                       \Delta \tau_{\kappa} \; \Delta \tau_{\kappa'}
$$
where $C$ is the interatomic force constant matrix and the sum on $\kappa$ and $\kappa'$ is
restricted to the Ti and O atoms that are allowed to move. With the help of this equation, we
can track the appearance of an instability in terms of the length of the chain of displaced
atoms. An instability will correspond to a specific displacement pattern that lowers the total
energy of the system: it will be associated to a negative eigenvalue of the restricted force
constant matrix.

In Fig.~\ref{Fig.chain}, we report the evolution of the lowest eigenvalue of the force constant
matrix with respect to the length of the chain of moving atoms. Displacing only a single atom,
the force induced on the Ti is larger than that on the O atom. With 3 atoms, we observe, at the
opposite, that the Ti-terminated chain (Ti--O--Ti) is more stable than the
O-terminated one (O--Ti--O): it points out the important role of the Ti--Ti interaction. The
difference between Ti and O terminated chains will disappear progressively with the
chain length. It is seen that an instability takes place for a chain longer than 10 atoms (5 unit
cells). This is in close agreement with the correlation length estimated in the previous
Section. It suggests that the behaviour of BaTiO$_3$ is already well reproduced when
considering an isolated Ti--O chain of displacements. It confirms that the correlation
between the different chains may play a minor role. 

Going further and frozing all the O atoms in such a way that only the Ti atoms are allowed to
move along the chain, we can repeat the previous investigations. For this case, however, we do
not observe any instability even for an infinite chain of correlated Ti displacements. This
result aims to prove that the relatively weak coupling between Ti and O displacements still
remains an important feature in the appearance of the structural instability.

\section{Conclusions}

In summary, we have reported first-principles linear response calculations of the phonon
dispersion curves of cubic BaTiO$_3$. Our results are in good agreement with the
experimental data.  Moreover, they are very similar to those reported for KNbO$_3$,
a cubic perovskite material presenting a sequence of phase transitions analogous to
BaTiO$_3$. In particular it was seen that a ferroelectric instability takes place when
correlated atomic displacements  are allowed along $<$100$>$ chains of finite length. This has
been investigated with a simplified model and the length of the shortest unstable chain
estimated to a dozen of atoms. The interatomic force constants are anisotropic and
propagate essentially along the Ti-O chains. Their analysis has emphasized the correlation of
the Ti displacements. It has shown that the Ti and O atomic displacements are only weakly
coupled. This small coupling remains however an important ingredient to reproduce the
ferroelectric instability.

\acknowledgments
We thank  J.-M. Beuken for permanent computer assistance. 
X.G. is grateful  to the FNRS-Belgium for financial support.
This paper presents research results of the Belgian Program on Interuniversity
Attraction Poles initiated by the Belgian State-Prime Minister's Office
Science Policy Programming. 
We aknowledge the use of the RS 6000 work
stations and the Namur Scientific Computing Facility (Namur-SCF),
which are common projects between IBM Belgium and respectively,
the Catholic University of Louvain (UCL-PCPM)
and the Facult\'es Universitaires Notre Dame de la Paix (FUNDP-LPMPS).

\begin{figure}
\caption{Convergence achieved on the calculated phonon dispersion curves of cubic
BaTiO$_3$ along the $\Gamma$-R line. The open symbols correspond to $q$-points included
in the M1 (circle)  and M2 (circle+square) meshes used to extrapolate the curves (M1: dotted
lines; M2: full lines). The filled symbols are associated to points not included in the mesh:
they illustrate that a satisfactory convergence is obtained with the M2 mesh.}
\label{Fig.GR}
\end{figure}

\begin{figure}
\caption{Calculated phonon dispersion curves of cubic BaTiO$_3$ at the experimental lattice
constant. The theoretical result shows a reasonable agreement with the experimental data:
($\bullet$) Ref. [3], ($\circ$) Ref. [6], ($+$) Ref. [7], ($\Box$) Ref. [8], ($\times$) Ref. [9],
($\nabla$) Ref. [10], ($\triangle$) Ref. [11].}
\label{Fig.BS}
\end{figure}

\begin{figure}
\caption{Zero-frequency isosurface of the lowest unstable phonon branch over the Brillouin
zone. $\Gamma$ is located at the center of the cube. The mode is unstable in the region
between the nearly flat surfaces.}
\label{Fig.IM}
\end{figure}

\begin{figure}
\caption{Lowest eigenvalue of the restricted force constant matrix associated to atomic
displacements along a finite Ti-O chain of increasing size.}
\label{Fig.chain}
\end{figure}

\begin{table}
\caption {Born effective charges and optical dielectric constant of
cubic BaTiO$_3$ at the experimental volume [27].}
\begin {tabular}{lccc}
   &Experiment    &Present   &Zhong {et al.}$^c$ \\
\hline
$Z^{*}_{Ba}$                         &$+2.9^a$     &$+2.74$       &$+2.75$        \\
$Z^{*}_{Ti} $                         &$+6.7^a$     &$+7.32$      &$+7.16$        \\
$Z^{*}_{O_{\perp}}$            &$-2.4^a$     &$-2.14$       &$-2.11$     \\
$Z^{*}_{O_{\parallel}}$       &$-4.8^a$     &$-5.78$      &$-5.69$        \\
$\epsilon_{\infty}$             &$5.40^b$     &$6.75$         &$-$ \\
\end{tabular}
$^a$ Reference \cite{Axe67} \\
$^b$ Reference \cite{Burns82} \\
$^c$ Reference  \cite{Zhong94}
\label{Table.Z}
\end{table}

\begin{table}
\caption{Computed phonon frequencies (cm$^{-1}$)of cubic BaTiO$_3$ at $\Gamma$, X, M and R. 
Symmetry labels follow the convention of Ref. [33]. }
\begin {tabular}{lcccccc}
$q$-point   &  &label   &frequency  & 
&label   &frequency\\
\hline
$\Gamma$  
 &   &$\Gamma_{15}$ (TO) &219 $i$ &   &     $\Gamma_{25}$            & 281 \\
&   &$\Gamma_{15}$ (A) & 0  &  &     $\Gamma_{15}$ (LO) & 445\\
&    &$\Gamma_{15}$ (LO) & 159 &  &     $\Gamma_{15}$ (TO) &453\\
&   &$\Gamma_{15}$ (TO) &166 &  &     $\Gamma_{15}$ (LO) & 631\\
\hline
X  & &X$_{5}$ &189 $i$ &  &     X$_{3}$ & 322\\
&   & X$_{5'}$ &104  &  &     X$_{5'}$ & 330\\
&   & X$_{2'}$ & 146 &  &     X$_{5}$ & 421\\
&   & X$_{5}$ & 194 &  &     X$_{1}$ & 517\\
&   & X$_{1}$ & 260  &  &     X$_{2'}$ & 627\\
\hline
M  & &M$_{3'}$ &167 $i$ &  &      M$_{5}$ &344\\
&   &  M$_{2'}$ &103  &  &      M$_{2}$ &354\\
&   & M$_{5'}$ &104  &  &      M$_{5'}$ &435\\
&    &  M$_{3}$ &208  & &      M$_{1}$ &456\\
&   &  M$_{5'}$ &270  &  &      M$_{4}$ &683\\
&   & M$_{3'}$ &333  &  \\
\hline
R &   &R$_{15}$ &128 &  &      R$_{25'}$ &386\\
&   &  R$_{25}$ &182  &  &      R$_{15}$ &414\\
&   & R$_{12'}$ &314  &  &      R$_{2'}$ &717\\
\end{tabular}
\label{Table.w}
\end{table}

\begin{table}
\caption {Longitudinal ($\parallel$) and transverse ($\perp$) interatomic force constants
(Ha/Bohr$^2$) with respect to a reference Ti atom (Ti(0)) along the Ti-O chain of cubic
BaTiO$_3$.}
\begin {tabular}{lccc}
Atom   &Total force    &DD force   &SR force \\
\hline
$Ti(0)$             &$+0.15215$     &$-0.27543$       &$+0.42758$        \\
\hline
$O_{\parallel}(1) $             &$+0.00937$     &$+0.23247$      &$-0.22310$        \\
$Ti_{\parallel}(2)$             &$-0.06721$     &$-0.03680$       &$-0.03041$     \\
$O_{\parallel}(3)$              &$+0.01560$     &$+0.00861$      &$+0.00699$        \\
$Ti_{\parallel}(4)$             &$-0.00589$     &$-0.00460$       &$-0.00129$      \\
\hline
$O_{\perp}(1) $             &$-0.02114$     &$-0.04298$       &$+0.02184$        \\
$Ti_{\perp}(2)$             &$+0.00751$     &$+0.01840$       &$-0.01089$     \\
\end{tabular}
\label{Table.IFC}
\end{table}

\begin{table}
\caption {Ti-Ti longitudinal interatomic force constants (Ha/Bohr$^2$) with respect to a
reference Ti atom at (.5, .5, .5).}
\begin {tabular}{lcccc}
coordinate   &distance  &IFC    &DD part   &SR part \\
\hline
(.5, .5, .5)         &0.0000   &$+0.15215$     &$-0.27543$       &$+0.42758$   \\
\hline
(-.5, .5, .5)       &7.5589   &-0.06721   &-0.03680   &-0.03041    \\
(-.5, -.5, .5)     &10.6899   &-0.01114   &-0.01301   &+0.00187    \\
(-.5, -.5, -.5)   &13.0924   &-0.00643   &-0.00780   &+0.00065    \\
(-1.5, .5, .5)     &15.1178   &-0.00589   &-0.00460   &-0.00129    \\
\end{tabular}
\label{Table.Ti-Ti}
\end{table}

\begin{table}
\caption {O-O longitudinal interatomic force constants (Ha/Bohr$^2$) with respect to a
reference O atom at (.5, .5, .0).}
\begin {tabular}{lcccc}
coordinate   &distance  &IFC    &DD part   &SR part \\
\hline
(.5, .5, .0)       &0.0000   &+0.12741   &-0.35322   &+0.48062   \\
\hline
(.5, .0, .5)       &5.3450   &-0.02838   &-0.03367   &+0.00529    \\
(-.5, .5, .0)     &7.5589   &-0.00190   &-0.00314   &+0.00124    \\
(.5, .5, -1.0)   &7.5589   &-0.03212   &-0.02295   &-0.00918    \\
(-.5, .0, .5)     &9.2577   &-0.00183   &-0.00289   &+0.00106    \\
(-.5,-.5, .0)    &10.6899   &-0.00290   &-0.00111   &-0.00179    \\
(-.5, .5, -1)    &10.6899   &-0.00415   &-0.00340   &-0.00078    \\
(.5, -1, -.5)    &11.9517   &-0.00254   &-0.00246   &-0.00008    \\
(-.5, -.5, -1)   &13.0924   &-0.00113   &-0.00129   &+0.00016    \\
\end{tabular}
\label{Table.O-O}
\end{table}


\begin{thebibliography}{10}
\bibitem{Comment2}
The values presented here are slightly different from those reported 
by us elsewhere  \cite{Ghosez94,Ghosez96+,Ghosez95+} due to the 
different convergence reached in the calculation.
\bibitem{Comment1}
The eigendisplacement vector $\eta$ was normalized such that $< \eta | M | \eta>=1$, where M
is such that $M=M_{\kappa} \; \delta_{\kappa, \kappa'}$ and $M_{\kappa}$ is the mass of atom
$\kappa$ in atomic mass units.
\bibitem{Comment3}
The length of the shortest unstable chain is slightly different from that reported for
KNbO$_3$. Changes in material properties could explain this difference although part of it
could be due to the different $k$-point and $q$-point convergence achieved in
Ref.~\cite{Yu95}: as observed in Fig.~\ref{Fig.GR}, the use of a finer mesh of $q$-points could
still slightly decrease the size of the zone of instability of BaTiO$_3$.
\end{thebibliography}
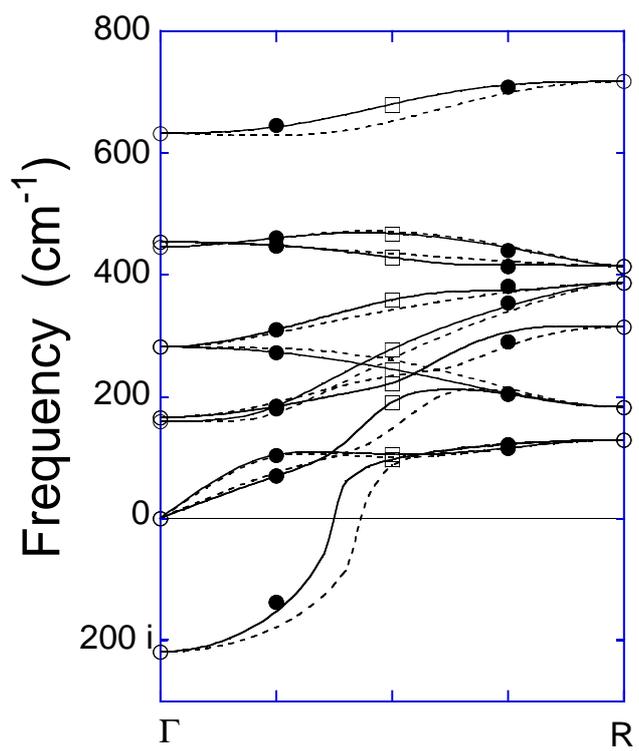
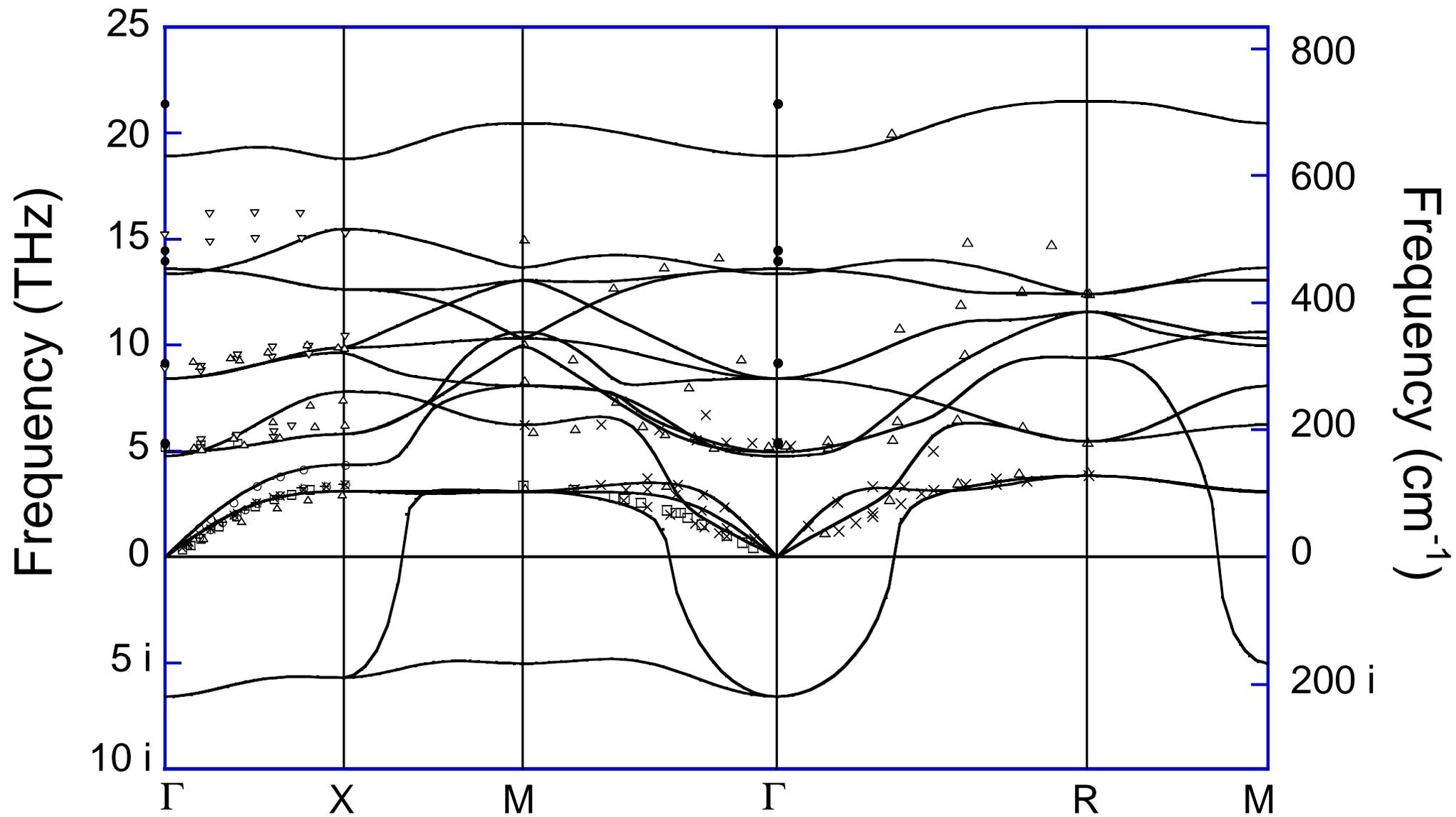
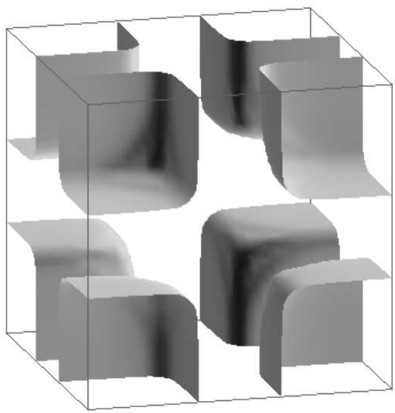
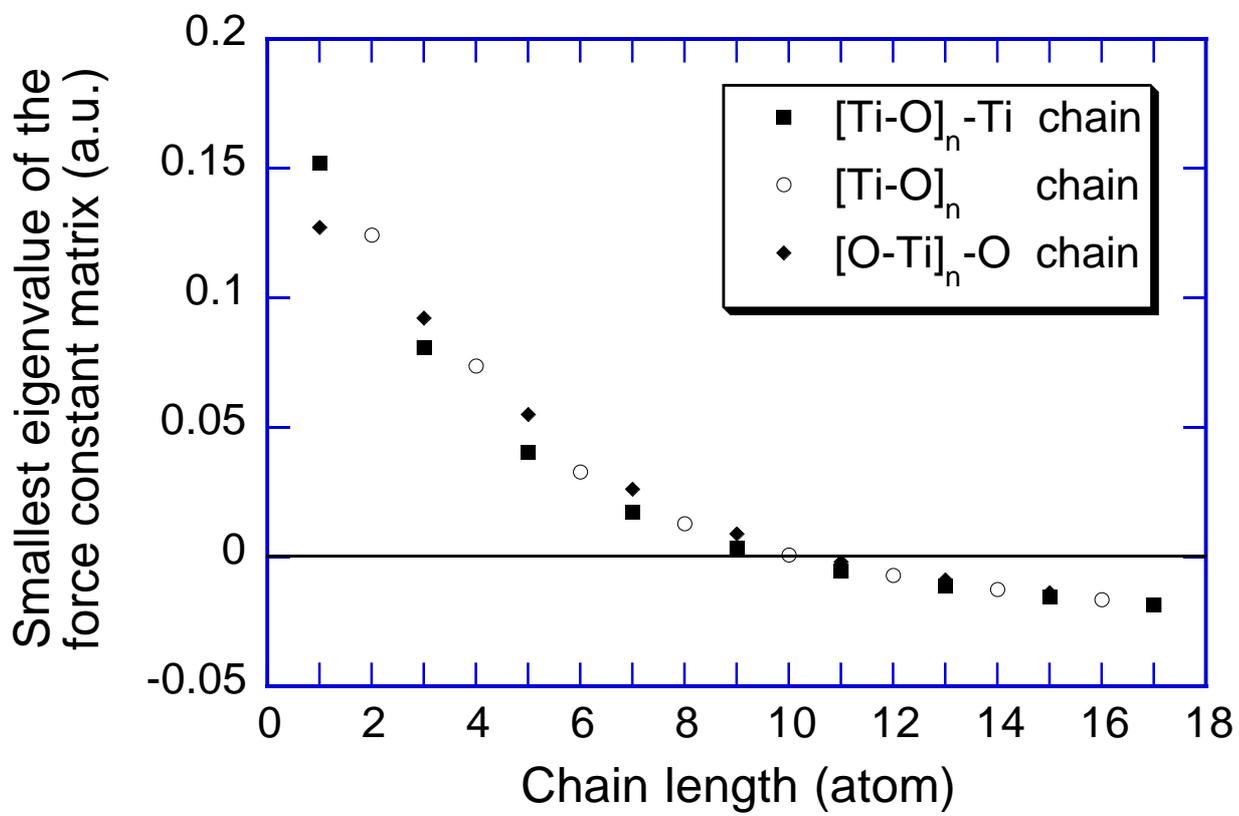
\end{document}